\documentclass[aps,prb,twocolumn,floats]{revtex4}

\usepackage{amssymb}
\usepackage{amsmath}
\usepackage{graphicx}
\usepackage{bm}
\usepackage{mdframed}
\usepackage[pangram]{blindtext}

\begin{document}

\bibliographystyle{prsty}
\author{H\'{e}ctor Ochoa,$^{1,2}$ Ricardo Zarzuela,$^{1,3}$ and Yaroslav Tserkovnyak$^{1}$}
\affiliation{$^{1}$ Department of Physics and Astronomy, University of California, Los Angeles, California 90095, USA\\
$^{2}$ Department of Physics, Columbia University, New York, NY 10027, USA\\
$^{3}$ Institut f\"{u}r Physik, Johannes Gutenberg Universit\"{a}t Mainz, D-55099 Mainz, Germany}

\begin{abstract}
We present a phenomenological theory of spin-orbit torques in a metallic ferromagnet with spin-relaxing boundaries. The model is rooted in the coupled diffusion of charge and spin in the bulk of the ferromagnet, where we account for the anomalous Hall effects as well as the anisotropic magnetoresistance in the corresponding constitutive relations for both charge and spin sectors. The diffusion equations are supplemented with suitable boundary conditions reflecting the spin-sink capacity of the environment. In inversion-asymmetric heterostructures, the uncompensated spin accumulation exerts a dissipative torque on the order parameter, giving rise to a current-dependent linewidth in the ferromagnetic resonance with a characteristic angular dependence. We compare our model to recent spin-torque ferromagnetic resonance measurements, illustrating how rich self-induced spin-torque phenomenology can arise even in simple magnetic structures.
\end{abstract}

\title{Self-induced spin-orbit torques in metallic ferromagnets}

\maketitle

\section{Introduction}

Spin-transfer torques in magnetic devices, i.e., the transfer of angular momentum leveraged by itinerant electrons to the magnetization dynamics,\cite{Berger,Slonczewski} enable the electrical control of the latter\cite{exp1,exp2,exp4,exp3} and are of interest for diverse technological applications. For example, spin-transfer torques can compensate the action of damping forces, sustaining a large-angle precessional motion in magnetic nano-oscillators,\cite{nano1,nano2,nano3} a phenomenon of potential interest in the field of neuromorphic computing.\cite{neuromorphic} Recent advances on this front exploit the torques of relativistic origin generated at the interface between a magnet and a heavy metal when charge flows in the latter.\cite{Scott} These torques can be described in terms of nonequilibrium accumulations due to the interfacial Edelstein \cite{Edelstein} and/or spin Hall effects,\cite{sH1,sH2} which rely on the lack of inversion symmetry imposed by the geometry of the device.

Spin accumulations can also be generated by spin-polarized currents in metallic ferromagnets without the active intervention of adjacent normal metals. In fact, the broken symmetries associated with the spontaneous magnetic ordering allow for more complex spin-current responses,\cite{spin-transfer1,spin-transfer2,spin-transfer3,perspective} such as the anisotropic magnetoresistance\cite{anisotropy} (AMR, which includes the so-called planar Hall effect\cite{pH}), leading to different mechanisms of spin transfer.\cite{SOT1,SOT2,SOT3} In this article, we present a minimal model for the spin-orbit torques generated by electronic currents in a heterostructure consisting of a ferromagnetic metal (FM) sandwiched between spin-relaxing layers, such as heavy normal metals (NM). Our theory is complementary to recent \textit{ab initio} studies.\cite{ab-initio1,ab-initio3,ab-initio2} The model relies on a phenomenological description of the flows of charge and longitudinal (to the magnetic order) spin, accompanied by suitable boundary conditions defined at the interfaces. We find that, when the heterostructure is inversion asymmetric, the uncompensated spin accumulation induced by a current density $\mathbf{j}_{\textrm{c}}$ exerts a damping-like torque (normalized by volume) on the magnetization of the form 
\begin{align}
\label{eq:anti-damping}
\boldsymbol{\tau}_{\textrm{d}}= & \frac{\eta \hbar}{2e\Gamma_s L}
\left(\bm{\hat{z}}\cdot\bm{n}\right)\,\left(\bm{n}\bm{\times} \bm{\hat{z}}\times \bm{n}\right)\\
&\times \left[\vartheta_{\textrm{H}}\,\bm{n}\cdot\left(\bm{\hat{z}}\bm{\times}\mathbf{j}_{c}\right)+ \varrho_{\textrm{MR}}\,\left(\bm{\hat{z}}\cdot\bm{n}\right)\left(\mathbf{j}_{c}\bm{\cdot}\bm{n}\right)\right].
\nonumber
\end{align}
Here, $\boldsymbol{n}$ is a unit vector along the collective spin density, $-e$ is the electron charge, $L$ is the thickness of the film, and $\Gamma_s$ is a dimensionless number characterizing the spin relaxation rate in the bulk of the ferromagnetic metal. The dimensionless coefficients $\vartheta_{\textrm{H}}$ and $\varrho_{\textrm{MR}}$ are related to the anomalous Hall and AMR effects, respectively, while $\eta$ (with units of the inverse volume) characterizes the torque by the out-of-equilibrium longitudinal spins on the order parameter at the interface. This spin torque can be either direct or mediated by magnons,\cite{Benedetta} leading to a characteristic temperature dependence in the latter case.\cite{tvetenPRB15} The above expression has been derived under the assumption that the magnetization dynamics occur on long timescales (longer than, e.g., the longitudinal spin-flip time, $\tau_s$), so that the itinerant electrons respond to a (quasi-)static magnetic background during their transport.

The torque~\eqref{eq:anti-damping} affects the linewidth of the ferromagnetic resonance as electron charge flows through the system. When the static component of the magnetization lies within the plane defined by the normal of the layered heterostructure and the charge current (the $xz$ plane in Fig.~\ref{fig:Fig1}), the shift in the resonance linewidth follows
\begin{align}
\Delta B\left(\theta\right)\approx\mathcal{A}_{xz}\Big(\sin2\theta+\textstyle{\frac{1}{2}}\sin4\theta\Big),
\end{align}
where $\theta$ is the polar angle of the magnetization relative to the $z$ axis and $\mathcal{A}_{xz}$ is the single fitting parameter proportional to $\varrho_{\textrm{MR}}/L$, see Eq.~\eqref{linewidth_PHE}. If, on the other hand, the order parameter lies within the plane perpendicular to the current and the heterostructure (the $yz$ plane in Fig.~\ref{fig:Fig1}), the shift in the resonance linewidth reads
\begin{align}
\label{eq:yz}
\Delta B\left(\theta\right)\approx\mathcal{A}_{yz}\left(\sin\theta+\sin3\theta\right),
\end{align}
where the prefactor $\mathcal{A}_{yz}$ now scales linearly with $\vartheta_{\textrm{H}}/L$. The linewidth shift vanishes for the magnetization within the film's ($xy$) plane, according to the torque \eqref{eq:anti-damping}. We note that these dependences for the resonance linewidth have been derived under the simplified assumption of circular precession of the magnetization. 

The manuscript is structured as follows: We present our model and derive the main results in Secs.~\ref{sec:theory} and~\ref{sec:linewidth}. In Sec.~\ref{sec:experiment}, we analyze data from recent spin-torque ferromagnetic resonance (ST-FMR) measurements\cite{SOT2} performed in different NM/FM heterostructures. 
We conclude by discussing and summarizing our findings in Sec.~\ref{sec:discussion}.

\section{Phenomenological model}

\label{sec:theory}

\begin{figure}[t!]
\includegraphics[width=\columnwidth]{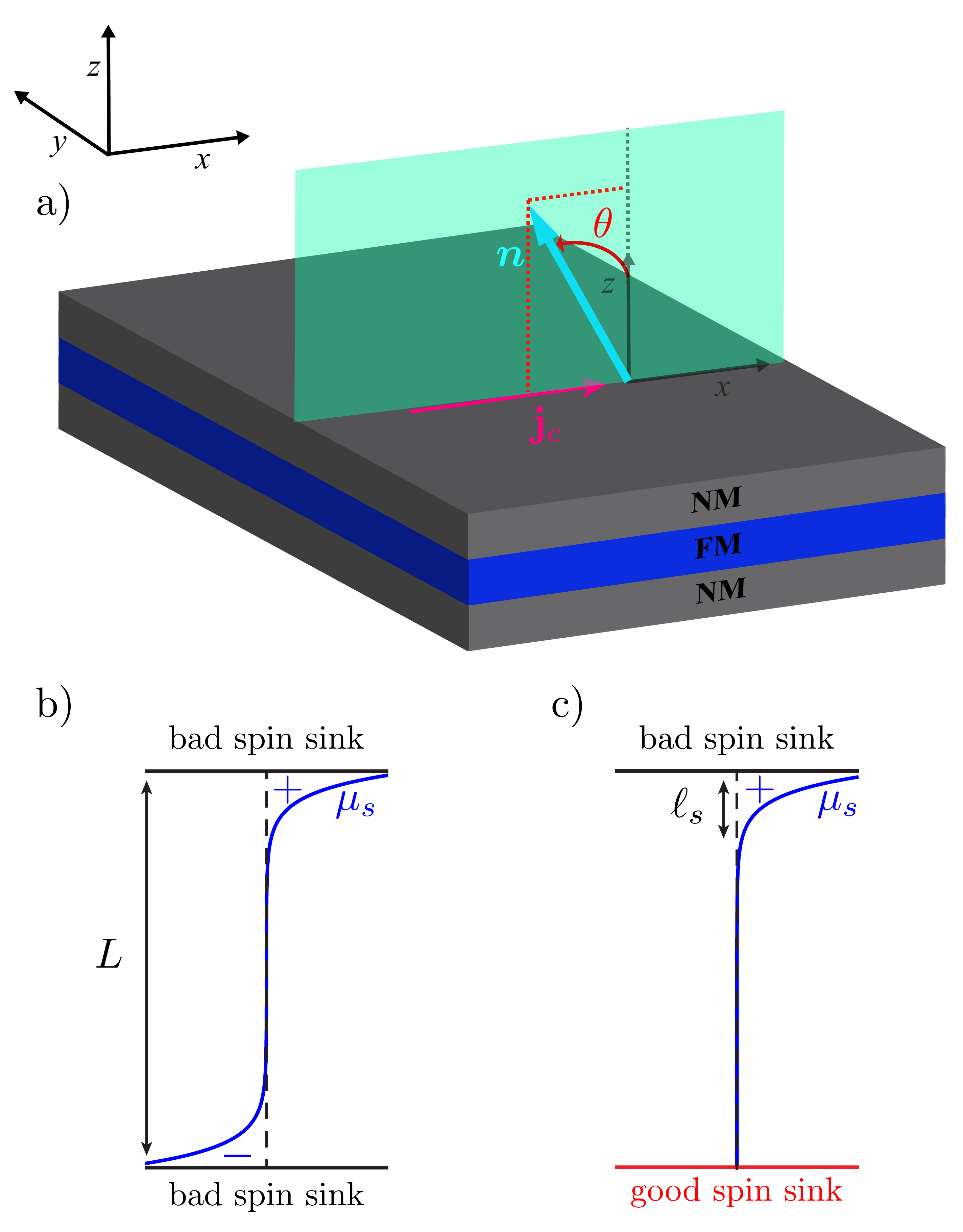}
\caption{a) Schematic representation of the heterostructure under consideration. b) Vertical profile of the spin accumulation within the ferromagnetic film (of thickness $L$) when both normal metals behave as bad spin sinks. c) The same when one of the metals is a good spin sink. The uncompensated spin accumulation is localized close to the interface with the bad spin sink on a length scale set by the spin diffusion length in the ferromagnet, $\ell_s$.}
\label{fig:Fig1}
\end{figure}

We consider a two-dimensional stack of a ferromagnetic conductor sandwiched between normal metals, see Fig.~\ref{fig:Fig1}(a). The relevant hydrodynamical variables are the charge density, $\rho_c$, and the longitudinal spin density, $\rho_s\equiv\boldsymbol{n}\cdot\boldsymbol{\rho}_s$. 
The conjugate thermodynamic forces are $\mu_c\equiv-e\,\delta_{\rho_c}\mathcal{F}$ and $\mu_s\equiv\hbar\, \delta_{\rho_s}\mathcal{F}$, where 
$\mathcal{F}$ is the free energy of the itinerant magnet. We assume that there are no slow variables related to transverse spin dynamics, apart from a coherent Landau-Lifshitz-type precession. In particular, any electronic spin dynamics relative to the collective order should relax very fast. In essence, we are constructing a phenomenology in which the spin degrees of freedom are coarse-grained down to the directional spin-density variable $\bm{n}$ and its magnitude that is parametrized by $\rho_s$ (which, while generally consisting of both the electronic and thermally-excited magnonic contributions, has our focus on the former). In the bulk of the ferromagnetic metal, we have local conservation laws of the form\begin{subequations}\begin{align}
& \partial_t\rho_c+\boldsymbol{\nabla}\cdot\mathbf{j}_c=0,\\
& \partial_t\rho_s+\boldsymbol{\nabla}\cdot\mathbf{j}_s=-\Gamma_{s}\,\mu_s,
\end{align}\end{subequations}
where $\Gamma_{s}=\hbar\nu_F/2\tau_s$. Here, $\tau_s$ and $\nu_F$ are, respectively, the spin-relaxation time and density of states per volume at the Fermi level. These continuity equations must be supplemented with constitutive relations of the form:
\begin{align}
\left(\begin{array}{c}
\mathbf{j}_c\\
\frac{2e}{\hbar}\mathbf{j}_s
\end{array}\right)=
\sigma\left(\begin{array}{cc}
\hat{\sigma}_c\left[\boldsymbol{n}\right] & \hat{\sigma}_{\textrm{x}}\left[\boldsymbol{n}\right] \\
\hat{\sigma}_{\textrm{x}}^T\left[-\boldsymbol{n}\right] & \hat{\sigma}_s\left[\boldsymbol{n}\right]
\end{array}\right)
\left(\begin{array}{c}
\frac{1}{e}\boldsymbol{\nabla}\mu_c\\
-\frac{1}{2e}\boldsymbol{\nabla}\mu_s
\end{array}\right),
\end{align}
where the off-diagonal matrix elements are related by the Onsager reciprocal relations.

In a featureless, isotropic ferromagnet, the normalized conductivity tensors have the following general structure: \begin{subequations}
\begin{align}
& \left[\hat{\sigma}_{c}\right]_{ij}=\delta_{ij}+\vartheta\,\epsilon_{ijk}n_k+\varrho\, n_in_j,\\
& \left[\hat{\sigma}_{s}\right]_{ij}=\delta_{ij}+\vartheta_{s} \,\epsilon_{ijk}n_k+\varrho_{s} \, n_in_j,\\
& \left[\hat{\sigma}_{\textrm{x}}\right]_{ij}=P\, \delta_{ij}+\vartheta_{\textrm{x}}\,\epsilon_{ijk}n_k+\varrho_{\textrm{x}}\, n_in_j.
\end{align}
\end{subequations}
Here, $\sigma$ is the total conductivity, which, in the two-channel phenomenology with little spin mixing, is given by $\sigma\approx\sigma_{\uparrow}+\sigma_{\downarrow}$, in terms of the conductivity $\sigma_{\uparrow}$ ($\sigma_{\downarrow}$)  of the majority (minority) electrons. The dimensionless parameter $P\approx(\sigma_{\uparrow}-\sigma_{\downarrow})/(\sigma_{\uparrow}+\sigma_{\downarrow})$ measures the spin polarization of the electrical current. The dimensionless coefficients $\vartheta$ and $\varrho$ parametrize the anomalous Hall and AMR effects, respectively. 
The coefficients $\vartheta_s$ and $\varrho_s$ parametrize analogous effects in the spin sector, while $\vartheta_{\textrm{x}}$ and $\varrho_{\textrm{x}}$ are associated with similar spin-charge cross terms. 
Microscopically, all these phenomenological constants depend on relativistic interactions and can typically be assumed to be small: $|\vartheta|\sim|\vartheta_{s,x}|,\varrho|\sim|\varrho_{s,x}|\ll 1$.

We apply these equations to the device geometry depicted in Fig.~\ref{fig:Fig1}(a), with the charge flowing in the $x$ direction, $\mathbf{j}_c=j\,\boldsymbol{\hat{x}}$. For simplicity, we assume translational invariance along $y$, and we focus on the charge and spin accumulations deep inside the ferromagnet, namely, far away from the leads.\cite{foot3} For the asymmetric case, consisting of the top/bottom normal metals being bad/good spin sinks, the spin accumulation along the transverse direction reads (see Appendix~\ref{AppA})
\begin{align}
\label{eq:mus}
\mu_s(z)=& \frac{2\,eE\,\ell_s'\sinh\left(\frac{z}{\ell_s'}\right)}{\left[1+\varrho_sn_z^2-\frac{\left(P+\varrho_{\textrm{x}} n_z^2\right)^2}{1+\varrho\,n_z^2}\right]\cosh\left(\frac{L}{\ell_s'}\right)}
\\
& \times \left[
\vartheta_{\textrm{x}}n_y+\varrho_{\textrm{x}}n_xn_z-
\frac{P+\varrho_xn_z^2}{1+\varrho\,n_z^2}\left(\varrho\,n_xn_z+\vartheta\,n_y\right)\right],
\nonumber
\end{align}
whose profile is shown in Fig.~\ref{fig:Fig1}(c). Here, $z>0$ is measured from the good spin sink and $\ell_{s}'$ denotes a magnetic-order dependent spin diffusion length, see Eq.~\eqref{spin_dif_length}. The parameter $E$ is associated with the voltage drop between the leads. In our final expression, we neglect a small misalignment between the applied current and the electric field associated with the Hall/MR effects, and write simply $E\approx j/\sigma$.

\section{Torque-induced linewidth}

\label{sec:linewidth}
Next we focus on the absorption power in a usual ferromagnetic resonance (FMR) experiment. We assume hereafter the low-frequency regime for the ac field, namely $\omega T_{1}\ll1$; furthermore, we assume that the corresponding wavelength is much larger than the size of the sample and, therefore, the itinerant ferromagnet exhibits a uniform dynamic state. The dynamics of the order parameter is described by the Landau-Lifshitz-Gilbert (LLG) equation,\cite{LL,G}
\begin{equation}
\label{LLG}
s(1+\alpha\,\boldsymbol{n}\times)\dot{\boldsymbol{n}}=\boldsymbol{n}\times\boldsymbol{H}_{\textrm{eff}}+\boldsymbol{\tau},
\end{equation}
where $s$ is the saturated spin density, $\alpha$ denotes the Gilbert damping constant and $\boldsymbol{H}_{\textrm{eff}}=-\delta \mathcal{F}/\delta\boldsymbol{n}$ is the thermodynamic force conjugate to the order parameter. To simplify the analysis, 
we will disregard in what follows anisotropy terms. Consequently, the magnetic free energy contains nothing but the Zeeman energy, $\mathcal{F}=-\gamma\, s\,\boldsymbol{n}\cdot(\boldsymbol{B}_{0}+\boldsymbol{b})$, with $\gamma$ being the gyromagnetic ratio and $\bm{B}_{0}$, $\bm{b}(t)$ denoting the strong dc and weak ac components of the magnetic field, respectively.

When reflection symmetry along the heterostructure axis ($\boldsymbol{\hat{z}}$) is broken while retaining the axial ($C_{\infty v}$) symmetry, the most generic torques to the lowest order in the spin accumulation can be written as\cite{Benedetta} 
\begin{align}
\label{eq:torque}
\boldsymbol{\tau}=\eta'\mu_s\left(\boldsymbol{\hat{z}}\cdot\boldsymbol{n}\right)\boldsymbol{n}\times \boldsymbol{\hat{z}}+\eta\,\mu_s\left(\boldsymbol{\hat{z}}\cdot\boldsymbol{n}\right)\boldsymbol{n}\times \boldsymbol{\hat{z}}\times \boldsymbol{n}.
\end{align}
Here $\eta$, $\eta'$ are phenomenological constants (with units of inverse of volume). One can imagine two possible microscopic mechanisms for these torques. In one scenario, the electronic spin accumulation exerts directly a torque on the order parameter due to the inversion-symmetry breaking-induced spin-orbit coupling at the interface. Another possibility is an inelastic channel mediated by magnons: the electronic spin accumulation is first converted into a magnon chemical potential,\cite{tvetenPRB15} and the magnon cloud subsequently exerts a torque on the coherent spin dynamics.\cite{Benedetta} The damping torque, second term in Eq.~\eqref{eq:torque}, reduces to the expression in Eq.~\eqref{eq:anti-damping}, where the prefactor comes from the average spin accumulation across the film thickness, $\bar{\mu}_s=\frac{1}{L}\int_0^{L}dz\,\mu_s\left(z\right)$. This is only different from $0$ for asymmetric heterostructures; from Eq.~\eqref{eq:mus}, to the leading order in relativistic effects (i.e., assuming $L\gg \ell_s$), we have
\begin{equation}
\label{eq:integrated_mu}
\bar{\mu}_s\approx\frac{2e\,\ell_s^{2}}{\sigma L}\big[\vartheta_{\textrm{H}}\,\bm{n}\bm{\cdot}(\bm{\hat{z}}\bm{\times}\mathbf{j}_{\textrm{c}})+\varrho_{\textrm{MR}}\,(\bm{\hat{z}}\bm{\cdot}\bm{n})(\mathbf{j}_{\textrm{c}}\bm{\cdot}\bm{n})\big],
\end{equation}
where we have introduced $\vartheta_{\textrm{H}}=\vartheta_{\textrm{x}}-P\vartheta$ and $\varrho_{\textrm{MR}}=\varrho_{\textrm{x}}-P\varrho$. Note that the prefactor can be conveniently written as
\begin{align}
\frac{2e\ell_s^{2}}{\sigma L}=\frac{\tau_s}{e\nu_FL}=\frac{\hbar}{2e\Gamma_{s} L},
\end{align}
yielding the prefactor in Eq.~\eqref{eq:anti-damping}.

In the measurements discussed later in Sec.~\ref{sec:experiment}, the static magnetic field was in the ballpark of 0.1 T,\cite{SOT2} which translates into a Larmor frequency of $\omega_{L}=\gamma B_{0}\simeq 10$ GHz. Just like in Ref.~\onlinecite{SOT2}, we assume for simplicity that in equilibrium the order parameter follows the static component of the magnetic field, $\bm{n}_{0}\propto\bm{B}_0$. When $\boldsymbol{b}(t)=\boldsymbol{b}\,e^{-i\omega t}$ is switched on, the order parameter acquires a small transverse ac component, $\boldsymbol{n}(t)=\boldsymbol{n}_{0}+\boldsymbol{\zeta}(t)$. The absorption power (averaged over an oscillation period) is proportional to the imaginary part of the transverse component of the susceptibility tensor, which can be obtained from the solution of the linearized LLG equation for $\boldsymbol{\zeta}$,
 \begin{align}
\label{chi}
\chi_{t}\left(\omega\right)=\frac{\gamma(\omega_{L}-i\alpha\omega)}{(\omega_{L}-i\alpha\omega)^{2}-\left[\omega+i\frac{\eta\bar{\mu}_{s}}{s}(\hat{\bm{z}}\bm{\cdot}\bm{n}_{0})^{2}\right]^{2}}.
\end{align}
Close to the resonant frequency, $\omega\approx\omega_L$, the absorption power goes as\begin{subequations}\begin{align}
P\left(\omega\right)\propto \omega\, \text{Im}\chi_t(\omega)\approx\frac{\gamma\omega}{2}\frac{\Gamma}{\left(\omega-\omega_L\right)^2+\Gamma^2},
\end{align}
with the resonance linewidth as a function of $\bm{n}_0$ and $\mathbf{j}_{\textrm{c}}$ given by\begin{align}
\label{linewidth}
\Gamma=\alpha\,\omega_L+\frac{\eta\hbar}{2es\Gamma_{s} L}\left(\boldsymbol{\hat{z}}\bm{\cdot}\boldsymbol{n}_0\right)^2 & \left[ \vartheta_{\textrm{H}}\,\boldsymbol{n}_0\bm{\cdot}\left(\boldsymbol{\hat{z}}\bm{\times}\mathbf{j}_{\textrm{c}}\right)\right.\\
& \left. +\varrho_{\textrm{MR}}\,\left(\boldsymbol{\hat{z}}\bm{\cdot}\boldsymbol{n}_0\right)\left(\mathbf{j}_{\textrm{c}}\bm{\cdot}\bm{n}_0\right)\right].
\nonumber
\end{align}
\end{subequations}

\section{Comparison to ST-FMR data}

\label{sec:experiment}

The current-induced shift in the resonance linewidth corresponds roughly to $\Delta B\sim\Gamma/\gamma$ after subtracting the Gilbert damping contribution,
\begin{align}
\label{linewidth_PHE}
\frac{\Delta B}{j_{c}}\approx & \frac{\eta\hbar}{8e\gamma s\Gamma_{s} L}\Big[\vartheta_{\textrm{H}}\left(\sin\theta+\sin3\theta\right)\sin\phi\\
& \hspace{1.1cm}+\varrho_{\textrm{MR}}\left(\sin2\theta+\textstyle{\frac{1}{2}}\sin4\theta\right)\cos\phi\Big], \nonumber
\end{align}where $\phi$ and $\theta$ denote the azimuthal and polar angles of the magnetization measured with respect to the direction of the current and $\bm{\hat{z}}$, respectively.

\begin{figure}[t!]
\includegraphics[width=\columnwidth]{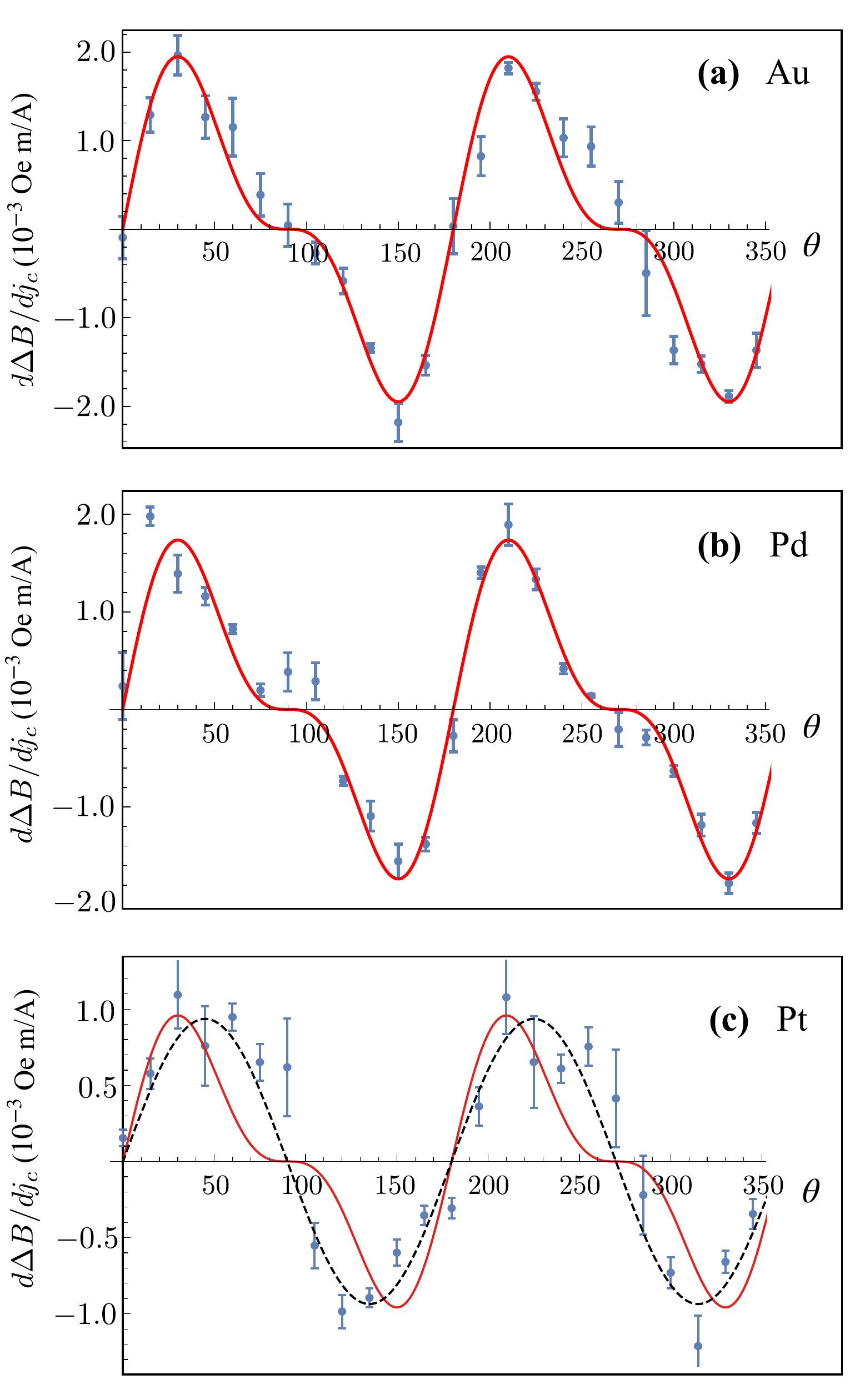}
\caption{Values of the resonance linewidth shift for the lowest-frequency mode measured in Ta/Au/FM/Ta [panel (a)], Ta/Pd/FM/Ta [panel (b)] and Ta/Pt/FM/Ta [panel (c)] at room temperature, extracted from Ref.~\onlinecite{SOT2}. Red lines represent the fitting of the second line of Eq.~\eqref{linewidth_PHE} to the data with $\mathcal{A}_{xz}=\eta\hbar\varrho_{\textrm{MR}}/8e\gamma s\Gamma_{s}L$. We obtain $\mathcal{A}_{xz}\simeq 1.5\cdot10^{-3}$ Oe m/A for Au and $\mathcal{A}_{xz}\simeq1.3\cdot10^{-3}$ Oe m/A for Pd, with coefficients of determination $R^{2}_{\textrm{Au}}=0.93$ and $R^{2}_{\textrm{Pd}}=0.95$, respectively. For Pt in panel (c) we obtain $\mathcal{A}_{xz}\simeq 0.74\cdot10^{-3}$ Oe m/A (with $R^{2}_{\textrm{Pt}}=0.68$). The blue dashed line represents the fitting of $\sin2\theta$ to the data, with $R^2=0.88$ in that case (the same fit to Au and Pd data yields similar values of $R^2$).}
\label{fig:Fig2}
\end{figure}

Figure~\ref{fig:Fig2} depicts the experimental values of the resonance linewidth shifts reported in Ref.~\onlinecite{SOT2} for three asymmetric nanostrips Ta/NM/FM/Ta, where FM denotes a [Co/Ni]$_{2}$/Co magnetic superlattice and NM is either Au [panel (a)], Pd [panel (b)] or Pt [panel (c)]. The corresponding measurements were taken at room temperature and FM layers of thickness $L=5.11$ nm were deposited during the fabrication of the nanostrips. In the same experiment, the current-induced shift was dramatically reduced for symmetric heterostructures.

The data points in Fig.~\ref{fig:Fig2}  corresponds to measurements with the static magnetic field lying within the plane defined by $\bm{\hat{z}}$ and the current ($\phi=0$), as depicted in the geometry of Fig.~\ref{fig:Fig1}. Red curves in Fig.~\ref{fig:Fig2} correspond to the second line of Eq.~\eqref{linewidth_PHE} with the overall factor $\eta\hbar/8e\gamma s\Gamma_{s} L$ and $\varrho_{\textrm{MR}}$ combined in a single fitting parameter, $\mathcal{A}_{xz}$. The formula reproduces well the angular dependence observed in the experiments for Au and Pd. In particular, our model captures the extra beating $\propto\sin 4\theta$ observed in the data, which goes beyond the behavior that may be naively expected for a magnetoresistance compatible with the reduced symmetry of the heterostructure, $(\bm{\hat{z}}\bm{\cdot}\bm{n})(\mathbf{j}_{c}\bm{\cdot}\bm{n})\propto\sin2\theta$. For Pt, our model does not yield the correct angular dependence for the resonance linewidth (neither does the fitting of the formula $\propto\sin2\theta$ showed in dashed line), which suggests that the strong spin-orbit interaction in Pt modifies physics beyond our simple spin-sink boundary condition.

The data for the case where the static field lies within the plane of the heterostructure ($\theta=\pi/2$) confirms this scenario (see Fig.~3a in Ref.~\onlinecite{SOT2}). Our model predicts no shift, which is indeed the case for Au within the experimental error, while for Pt heterostructures there is a sizable shift resulting from a more conventional mechanism rooted in the spin Hall effect (see Appendix~\ref{AppB}) originating in the heavy metal,\cite{exp4} which appears to be the largest contribution for the totality of the angular dependence ($\propto\sin\phi$ in this configuration, with some corrections). 

\section{Discussion}

\label{sec:discussion}

One important property of our model is that it reproduces well the extra beating in the angular dependence of the linewidth shifts of Au and Pd heterostructures. Specifically, the model fixes the relative strength of the $\sin2\theta$ and $\sin 4\theta$ components in the signal, with only one global fitting parameter measuring the overall strength of the current-induced shift. The similar values of $\mathcal{A}_{xz}$ for both nanostrips ($\mathcal{A}_{xz}\simeq 1.5\cdot10^{-3}$ Oe m/A for Au and $\mathcal{A}_{xz}\simeq1.3\cdot10^{-3}$ Oe m/A for Pd) agrees well with the basic ingredient of the model, namely, that the exact nature of the normal metals play a secondary role beyond defining the boundary conditions for the coupled spin-charge diffusion in the ferromagnetic metal. The interface with Pt, however, seems to play a more active role in the magnetization dynamics. The same trend is confirmed by the data in the $xy$ plane of Ref.~\onlinecite{SOT2}. Our model predict a nil shift, compatible with the data in Au heterostructures. There is, however, a sizable shift coming from the adjacent Pt film and also, to a smaller extent, in the Pd case. We conclude that the spin-orbit effects in Pt give rise to the more conventional external torques,\cite{Scott} while Au is dominated by our internal mechanism. Pd (which is electronically similar to Pt, but with a weaker spin-orbit interaction) seems to be somewhat intermediate and displays both mechanisms, with a stronger self-induced torque, as suggested by the good fit to the $xz$ plane data shown in Fig.~\ref{fig:Fig2}b.

The previous analysis and, in particular, the disagreement between our model and the experimental data for Pt makes clear that the present theory is not the most general one. Thus, it is worth discussing our results in the context of a more general phenomenology guided by symmetry, in the spirit of Ref.~\onlinecite{Garello_etal}. In order to make contact with our model, in the following construction, we incorporate the separation of time scales between the dynamics of the order parameter and the itinerant degrees of freedom, by considering only symmetry-allowed interfacial torques up to linear order in the current density $\mathbf{j}_{\textrm{c}}$. For simplicity, we discuss only asymmetric heterostructures with the principal axis oriented along $\bm{\hat{z}}$. As in Sec.~\ref{sec:theory}, the ferromagnets are assumed to be isotropic, while the presence of normal metals reduces the symmetry down to $C_{\infty v}$ (for the subsequent notation, see Ref.~\onlinecite{book}).

Torques must be orthogonal to the magnetic order $\bm{n}$, yielding two possibilities at the interface, up to  a (pseudo)scalar prefactor:
\begin{subequations}
\label{eq:vectors}
\begin{align}
& \bm{\hat{z}}\times\bm{n},\\
& \bm{n}\times\bm{\hat{z}}\times\bm{n}.
\label{eq:vector-PHE}
\end{align}
\end{subequations}
The torques must also transform as $\bm{n}$, namely, the $z$ component must be a pseudoscalar ($A_2$ representation), and the rest of the components form a vector ($E_1$ representation). The two candidates in Eqs.~\eqref{eq:vectors} behave accordingly only if multiplied by a pseudoscalar (e.g., $n_z$, which leads to Eq.~\ref{eq:torque}). Here, we consider all the possible pseudoscalars up to linear order in the current density. Decomposing $\mathbf{j}_{\textrm{c}}$ in its collinear and orthogonal components to the projection of $\bm{n}$, we have again two possibilities:\begin{subequations}
\label{eq:scalars}
\begin{align}
\label{eq:psedu-Hall}
& \bm{n}\cdot\mathbf{j}_{\textrm{c}},\\
& \left({\bm{\hat{z}}}\cdot\bm{n}\right)\left[\bm{n}\cdot\left(\bm{\hat{z}}\times\mathbf{j}_{\textrm{c}}\right)\right].
\label{eq:psedu-PHE}
\end{align}
\end{subequations}
The combination of these two pseudoscalars with the two vectors in Eq.~\eqref{eq:vectors} generate four groups of magnetic torques.\cite{book} In particular, the Hall torque, first term in Eq.~\eqref{eq:anti-damping}, follows directly by combining Eqs.~\eqref{eq:vector-PHE} and \eqref{eq:psedu-PHE}. Additional ST-FMR measurements with the static component of the field within the plane perpendicular to the current ($\phi=\pi/2$ in our expressions, $yz$ plane in Fig.~\ref{fig:Fig1}) would directly test this Hall contribution, as described by Eq.~\eqref{eq:yz} in our model.

In general, we should include higher powers in ${\bm{\hat{z}}}\cdot\bm{n}$ (only even powers are allowed by symmetry) weighted by different phenomenological constants; for example, \begin{align}
\hspace{-0.2cm} \left(\bm{n}\cdot\mathbf{j}_{\textrm{c}}\right)\left[A+B(\bm{\hat{z}\cdot\bm{n}})^2+...\right]\bm{n}\times\bm{\hat{z}}\times\bm{n}.
\end{align}
The magnetoresistance torque, second term in Eq.~\eqref{eq:anti-damping}, corresponds to the case of $A=0$, which is specific to our transport/torque model. For a more general expansion as in Ref.~\onlinecite{Garello_etal}, we should allow also for geometrical factors of the form $1/[1-(\bm{\hat{z}\cdot\bm{n}})^2]$, as is the case, for example, for the usual spin Hall torque generated by spin accumulation in an adjacent normal metal, see Appendix~\ref{AppB}.


In conclusion, we have presented a theory for self-induced torques in NM1/FM/NM2 heterostructures. The model relies on the separation of time scales between the magnetization and electron dynamics. The latter is described by diffusion equations for the charge and longitudinal spin coupled via constitutive relations that include the anomalous Hall and AMR effects in the bulk of the ferromagnet generated by the static component of the magnetization. Both effects produce steady-state spin accumulations at the interfaces with the normal metals which, in turn, exert a net torque on the order parameter if uncompensated. The damping-like interfacial torques are manifested through characteristic model-specific beatings in the angular dependence of the ST-FMR resonance linewidths. Other signatures are the dependence on the thickness of the heterostructure or the temperature dependence contained in $\tau_s$ and the coupling $\eta$.

\begin{acknowledgments}
The authors are grateful to Eric Montoya and Ilya Krivorotov for sharing their data before publication and bringing this problem to our attention. This work was supported by the U.S. Department of Energy, Office of Basic Energy Sciences under Award No.~DE-SC0012190.
\end{acknowledgments}

\appendix

\section{Spin accumulation in the steady state}

\label{AppA}

Translational invariance along $y$ yields that the electrochemical potential $\mu_c$ and the spin accumulation $\mu_s$ to be functions of the coordinates $x$ and $z$ only. Since we have restricted ourselves to a spatial region of the magnet far way from the leads, we also disregard the dependence of $\mu_s$ on the coordinate $x$ and approximate $\mu_c(x,z)\approx eE x+\tilde{\mu}_c(z)$. 
Supposing that the noncollinearity between the applied current and the induced electric field is small, we neglect it when writing Eq.~\eqref{eq:integrated_mu}, and approximate $E\approx j/\sigma$.

The spin and charge accumulations along $z$ are related by the condition $\boldsymbol{\hat{z}}\cdot\mathbf{j}_c=0$, that is
\begin{align}
\left(1+\varrho\, n_z^2\right)\partial_z\tilde{\mu}_c+\left(\vartheta\, n_y+\varrho\, n_xn_z\right)eE=\frac{P+\varrho_{\textrm{x}}n_z^2 }{2}\, \partial_z\mu_s.
\end{align}By using this last relation, the spin continuity can be recast as
\begin{align}
\left[1+\varrho_s\,n_z^2-\frac{\left(P+\varrho_{\textrm{x}} n_z^2\right)^2}{1+\varrho\,n_z^2}\right]\partial_z^2\mu_s=\frac{\mu_s}{\ell_s^2},
\end{align}where $\ell_s\equiv\sqrt{\hbar\sigma/4e^2\Gamma_{s}}$ is the spin diffusion length. The solution to this equation reads in general
\begin{align}
\mu_s\left(z\right)=\mu_b\,e^{-\frac{z}{\ell_s'}}+\mu_t\,e^{\frac{z-L}{\ell_s'}},
\end{align}
with
\begin{align}
\label{spin_dif_length}
\ell_s'=\ell_s\sqrt{1+\varrho_s\,n_z^2-\frac{\left(P+\varrho_{\textrm{x}} n_z^2\right)^2}{1+\varrho\,n_z^2}}.
\end{align}The coefficients $\mu_{b,t}$ can be inferred from the boundary conditions on the spin flow across the bottom/top interfaces, placed at $z=0$ and $z=L$, respectively.

\subsubsection{Both normal metals are poor spin sinks}

In that case the condition $\boldsymbol{\hat{z}}\cdot\mathbf{j}_s=0$ holds at both interfaces. The spin accumulation along the transverse direction reads then
\begin{align}
\mu_s(z)=& \frac{2\,eE\,\ell_s'\sinh\left(\frac{z-L/2}{\ell_s'}\right)}{\left[1+\varrho_sn_z^2-\frac{\left(P+\varrho_{\textrm{x}} n_z^2\right)^2}{1+\varrho\,n_z^2}\right]\cosh\left(\frac{L}{2\ell_s'}\right)}
\\
& \times \left[
\vartheta_{\textrm{x}}n_y+\varrho_{\textrm{x}}n_xn_z-
\frac{P+\varrho_xn_z^2}{1+\varrho\,n_z^2}\left(\varrho\,n_xn_z+\vartheta\,n_y\right)\right],
\nonumber
\end{align}whose profile is depicted in Fig.~\ref{fig:Fig1}(b).

\subsubsection{Good/poor spin sinks at the bottom/top interfaces}

Now we consider the asymmetric case where the top (bottom) normal metal is, as before, a bad (good) spin sink; we have then the identity $\mu_s(z=0)=0$, i.e., there is no spin accumulation at the interface with the good spin sink. The spin accumulation is given in this case by Eq.~\eqref{eq:mus}.

\section{Spin Hall torque}

\label{AppB}

An asymmetric heterostructure should generally yield also the spin Hall torque, which can be generated by a heavy normal metal such as Pt:\cite{Scott}\begin{align}
\label{eq:B1}
\bm{\tau}_{\textrm{sH}}=\left(\eta_{\textrm{sH}}+\vartheta_{\textrm{sH}}\,\bm{n}\times\right)\left(\bm{\hat{z}}\times\mathbf{j}_{\textrm{c}}\right)\times\bm{n}.
\end{align}
Here $\eta_{\textrm{sH}}$ is a phenomenological parameter associated, for example, with the Edelstein effect\cite{Edelstein} and parametrizing the reactive component of the torque; $\vartheta_{\textrm{sH}}$ is the spin-Hall angle characterizing the dissipative counterpart. Here we show that this torque can be expressed in terms of the basis introduced in Eqs.~\eqref{eq:vectors} following the prescription discussed in the main text.

First observe that we can write $(\bm{\hat{z}}\times\mathbf{j}_{\textrm{c}})\times\bm{n}=(\bm{\hat{z}}\cdot\bm{n})\,\mathbf{j}_{\textrm{c}}-(\bm{n}\cdot\mathbf{j}_{\textrm{c}})\,\bm{\hat{z}}$. By decomposing $\mathbf{j}_{\textrm{c}}$ in its longitudinal and transverse components to $\bm{n}$, we have then\begin{widetext}\begin{align}
(\bm{\hat{z}}\times\mathbf{j}_{\textrm{c}})\times\bm{n}=-(\bm{n}\cdot\mathbf{j}_{\textrm{c}})\left(\bm{n}\times\bm{\hat{z}}\times\bm{n}\right)+(\bm{\hat{z}}\cdot\bm{n})\left(\bm{n}\times\mathbf{j}_{\textrm{c}}\times\bm{n}\right).
\end{align}
The first term is already of the desired form, namely, a combination of vector~\eqref{eq:vector-PHE} and pseudoscalar~\eqref{eq:psedu-Hall}. For the second term, the transverse component of the current can be expanded in the basis of Eqs.~\eqref{eq:vectors}; specifically, we have\begin{align}
\label{eq:B3}
\bm{n}\times\mathbf{j}_{\textrm{c}}\times\bm{n}=\frac{\bm{n}\cdot\left(\bm{\hat{z}}\times\mathbf{j}_{\textrm{c}}\right)}{(\bm{\hat{z}}\cdot\bm{n})^2-1}\,\bm{\hat{z}}\times\bm{n}+\frac{(\bm{\hat{z}}\cdot\bm{n})(\bm{n}\cdot\mathbf{j}_{\textrm{c}})}{(\bm{\hat{z}}\cdot\bm{n})^2-1}\,\bm{n}\times\bm{\hat{z}}\times\bm{n}.
\end{align}
Regrouping all these terms, we can finally write\begin{align}
(\bm{\hat{z}}\times\mathbf{j}_{\textrm{c}})\times\bm{n}=\frac{1}{(\bm{\hat{z}}\cdot\bm{n})^2-1}\left[(\bm{n}\cdot\mathbf{j}_{\textrm{c}})\left(\bm{n}\times\bm{\hat{z}}\times\bm{n}\right)+\left({\bm{\hat{z}}}\cdot\bm{n}\right)\left[\bm{n}\cdot\left(\bm{\hat{z}}\times\mathbf{j}_{\textrm{c}}\right)\right]\left(\bm{\hat{z}}\times\bm{n}\right)\right].
\end{align}
\end{widetext}
The terms between brackets are of the desired form: a combination of vector~\eqref{eq:vector-PHE} and pseudoscalar~\eqref{eq:psedu-Hall} (first term) and the same vector with pseudoscalar~\eqref{eq:psedu-PHE} (second term). The overall prefactor is a scalar that seems to result in a divergence as $\bm{n}$ approaches $\bm{\hat{z}}$. This is just an artifact of the basis choice in Eqs.~\eqref{eq:vectors}, which is ill-defined in that limit. Note, however, that the overall expression in the right-hand side of this last equation is smooth and gives the correct limit $(\bm{\hat{z}}\times\mathbf{j}_{\textrm{c}})\times\bm{n}\rightarrow \mathbf{j}_{\textrm{c}}$ when $\bm{n}\rightarrow\bm{\hat{z}}$.

Finally, from Eq.~\eqref{eq:B3}, it also directly follows that the two other basic spin-orbit torques, $(\bm{\hat{z}}\cdot\bm{n})\,\mathbf{j}_{\textrm{c}}\times\bm{n}$ and $(\bm{\hat{z}}\cdot\bm{n})\,\bm{n}\times\mathbf{j}_{\textrm{c}}\times\bm{n}$ (which, together with torques \eqref{eq:B1}, constitute the full basis\cite{Garello_etal}) can be similarly expanded in terms of Eqs.~\eqref{eq:vectors}~and~\eqref{eq:scalars}.

\end{document}